\documentclass[aip,apl, reprint]{revtex4-1}

\usepackage{graphicx}
\usepackage{soul}
\usepackage{amsmath}
\usepackage[utf8]{inputenc}

\begin{document}


\title{Spin-dependent recombination at arsenic donors in ion-implanted silicon}


\author{David P. Franke}
\email{david.franke@wsi.tum.de}
\affiliation{Walter Schottky Institut and Physik-Department, Technische Universität München, Am Coulombwall 4, 
85748 Garching, Germany}
\author{Manabu Otsuka}
\affiliation{School of Fundamental Science and Technology, Keio University, 3-14-1 Hiyoshi, 
Kohoku-ku, Yokohama 223-8522, Japan}

\author{Takashi Matsuoka}
\affiliation{School of Fundamental Science and Technology, Keio University, 3-14-1 Hiyoshi, 
Kohoku-ku, Yokohama 223-8522, Japan}
\author{Leonid S. Vlasenko}
\author{Marina P. Vlasenko}
\affiliation{A.~F.~Ioffe Physico-Technical Institute, Russian Academy of Sciences, 
194021, St.~Petersburg, Russia}
\author{Martin S. Brandt}
\affiliation{Walter Schottky Institut and Physik-Department, Technische Universität München, Am Coulombwall 4, 
85748 Garching, Germany}
\author{Kohei M. Itoh}
\affiliation{School of Fundamental Science and Technology, Keio University, 3-14-1 Hiyoshi, 
Kohoku-ku, Yokohama 223-8522, Japan}

\date{\today}

\begin{abstract}
Spin-dependent transport processes in thin near-surface doping regions created by low energy ion implantation of arsenic in silicon are detected by two methods, spin-dependent recombination (SDR) using microwave photoconductivity and electrically detected magnetic resonance (EDMR) monitoring the DC current through the sample. The high sensitivity of these techniques allows the observation of the magnetic resonance in particular of As in weak magnetic fields and at low resonance frequencies (40--1200 MHz), where high-field-forbidden transitions between the magnetic substates can be observed due to the mixing of electron and nuclear spin states. Several implantation-induced defects are present in the samples studied and act as spin readout partner. We explicitly demonstrate this by electrically detected electron double resonance experiments and identify a pair recombination of close pairs formed by As donors and oxygen-vacancy centers in an excited triplet state (SL1) as the dominant spin-dependent process in As-implanted Czochralski-grown Si.
\end{abstract}


\maketitle

In the fabrication of semiconductor devices, ion implantation is widely used to create thin layers doped with different impurities for the realization of, e.g., $p$-$n$ junctions or ohmic contacts. Electron paramagnetic resonance (EPR) is the characterization method for the identification of the dopants and defects created during the implantation process and the investigation of their microscopic structure \cite{brower_electron_1970, barklie_electron_2005}. However, conventional EPR measurements can usually only be performed on samples implanted with relatively high ion energies and ion doses to get the necessary number of defects. 
One possibility to enhance the sensitivity of EPR measurements is the detection of spin-dependent conductivity (SDC) in methods such as electrically detected magnetic resonance (EDMR)\cite{schmidt_modulation_1966, stich_electrical_1995, stegner_electrical_2006} or spin-dependent recombination (SDR) \cite{vlasenko_electron_1995}. Indeed, using both experimental approaches, dopants and defects after low dose low energy implantation have been studied succesfully. SDR, where EPR spectra are detected by measuring the microwave reflectivity of the sample, has been applied to, e.g.,  the investigation of defects after implantation of hydrogen \cite{laiho_electron_1999} and bismuth \cite{mortemousque_spin_2012}.
In EDMR, where samples are equipped with electrical contacts to observe SDC by monitoring the DC photoconductivity, as few as 50 phosphorus donors could be detected, implanted at an energy of $14$ keV (Ref.~\onlinecite{mccamey_electrically_2006}). Using single electron transistors for detection, the electron and nuclear spin state even of single low-energy-implanted phosphorus donors in Si can be measured\cite{morello_single-shot_2010}.
In addition to the higher sensitivity when compared to conventional EPR, a specific feature of all SDC-based mechanisms is the weak dependence of the resonance line intensities on the strength of the external magnetic field \cite{vlasenko_spin-dependent_1986, brandt_electrically_1998, morishita_electrical_2009}. This allows to observe EPR spectra at weak magnetic fields and low resonance frequencies without a loss of sensitivity \cite{franke_spin-dependent_2014}.
A particular benefit of this is the possibility to extend EPR experiments into a regime where the hyperfine interaction between the electron and the nuclear spin is comparable in strength to the electronic Zeeman interaction, allowing to excite transitions between magnetic sublevels via EPR which are forbidden in higher magnetic fields \cite{morishita_electrical_2009}.

In this letter, we present a combined SDR and EDMR investigation into the spin-dependent recombination involving defects produced by low energy As$^+$ ion implantation in silicon. Resonance line positions at low magnetic fields ($\leq 50$ mT) are investigated by SDR and behave as expected regarding the mixing of electron and nuclear spin states.
Furthermore, the spin-dependent processes leading to SDC signals for As donors and defects in an excited triplet state (SL1) after implantation in oxygen-rich silicon are investigated with pulsed EDMR methods and evidence for a spin-pair mechanism is found.

As$^+$ ions with energies of 30 keV and doses between $10^{13}$ and $5\cdot 10^{15}$ cm$^{-2}$ were implanted in high resistance ($>300~\Omega$cm, $n$-type) wafers of float-zone (FZ) and Czochralski (Cz) grown silicon. SDR measurements were performed on samples $\sim 8\times 4 \times 0.3$ mm$^3$ in size with an X-band spectrometer (JEOL JES-RE3X) with a cylindrical TE$_{011}$ cavity and magnetic field modulation with a modulation frequency of 100 kHz. The SDR signal was detected as a change in microwave photoconductivity of the sample illuminated by white light from a $100$ W halogen lamp under irradiation of 100--200 mW of microwave (mw) saturating the EPR transition.
The temperature $T$ could be varied in the range of 4--35 K using a helium gas flow cryostat. To suppress the background due to the magnetoresistance of the sample, the second derivative of the SDR signal was recorded.

\begin{figure}%
\includegraphics[width=\columnwidth]{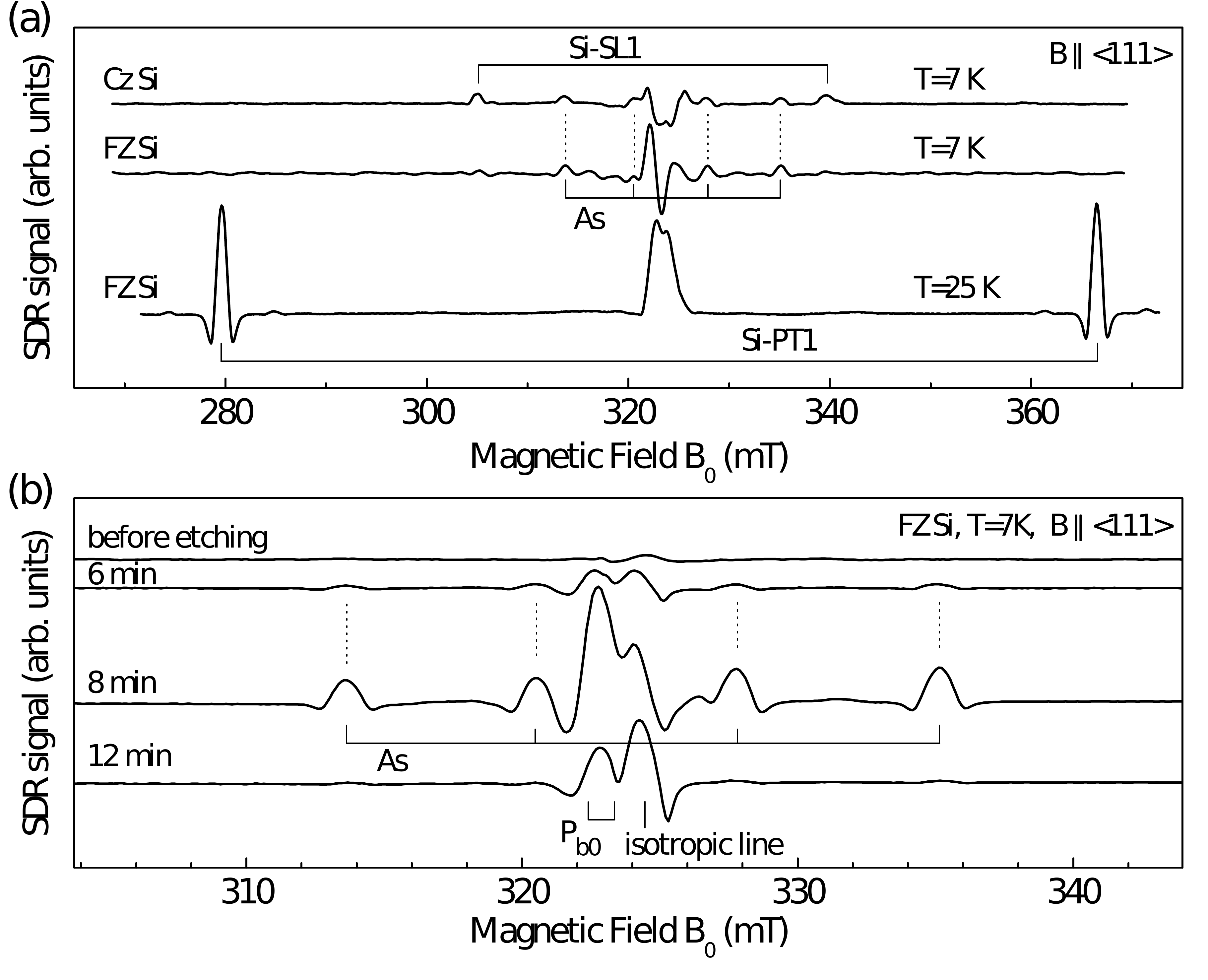}%
\caption{(a) High-field SDR spectra of Cz and FZ Si after 30 keV As$^+$-ion implantation (dose $5\cdot 10^{15}$ cm$^{-2}$) as detected before annealing. (b) High-field SDR spectra of the FZ Si sample after subsequent annealing for 30 min at $750^{\circ}$C and etching in KOH for different etching times.}%
\label{figSDRspec}%
\end{figure}

Before ion implantation, the spectra of surface recombination centers known as P$_\mathrm{b0}$ centers \cite{poindexter_interface_1981, stesmans_electron_1998} and an isotropic line with a $g$-factor of about $1.9996$ (possibly due to conduction-band-related paramagnetic species, cf.~Ref.~\onlinecite{graeff_electrically_1999, Lu_high-resolution_2011, matsuoka_identification_2012}) were observed in SDR for all samples. 
After implantation, the spectra of radiation defects were observed as shown in Fig.~\ref{figSDRspec} (a). In oxygen-rich Cz Si, the SL1 spectrum \cite{brower_electron_1971} belonging to the excited triplet state of the neutral oxygen-vacancy center OV$^0$ is found. For the FZ sample, the PT1 signal arising from the excited triplet state of a complex formed by two substitutional carbon atoms and one interstitial silicon atom (C$_\mathrm{s}$--Si$_\mathrm{i}$--C$_\mathrm{s}$)\cite{vlasenko_EPR_1984-1, vlasenko_electron_1995} is detected. Different measurement temperatures were chosen to optimize the signal-to-noise ratios and are given in Fig.~\ref{figSDRspec} (a). The intensities of these resonances increase with increasing implantation doses up to the highest dose studied here.
Weak SDR spectra of isolated As dopants are detected in the samples of FZ and Cz Si subjected to implantation doses higher than $5\cdot 10^{14}$ cm$^{-2}$.

After high-temperature annealing ($600$--$900^{\circ}$C) the SDR signal intensity was found to be strongly reduced, which can be explained by the formation of a low resistance layer near the sample surface reducing the quality factor of the mw cavity and, consequently, the sensitivity of the SDR detection. To remove this layer, the samples were etched in 15\% KOH in H$_2$O at room temperature, resulting in an etching rate of $15$--$20$ nm/min (Ref.~\onlinecite{seidel_anisotropic_1990}). The SDR spectra of FZ samples subjected to this etch for different times are shown in Fig.~\ref{figSDRspec} (b). For small etch durations, an increase in the intensity of all lines with etching time is observed. After 12 min, however, the lines corresponding to As donors have disappeared, in agreement with the expected implantation depth of $\sim70$ nm for an As$^+$-ion energy of $30$ keV.

One of the important specific features of SDC-based measurements is the weak dependence of the signal intensity on the strength of the external magnetic field. This is due to the fact that the detection of SDC does not depend the Boltzmann spin polarization, but is governed by the spin symmetry of close pairs, like weakly bound electron-hole pairs in amorphous semiconductors \cite{kaplan_explanation_1978} or donor-acceptor pairs in crystals \cite{cox_exchange_1978, stich_electrical_1995}. Therefore, it is possible to detect magnetic resonance signals at weak magnetic fields, where also transitions between magnetic sublevels which are forbidden in the high-field limit can be observed. For the excitation of magnetic resonance transitions at weak magnetic fields and  low resonance frequencies (40--1200 MHz), a wire was coiled around the FZ Si sample and connected to a radio-frequency (rf) oscillator. At the same time, the reflectivity of the microwave cavity (frequency $\sim 9$ GHz) was recorded to detect changes in microwave photoconductivity of the sample at rf excitation.

\begin{figure}%
\includegraphics[width=\columnwidth]{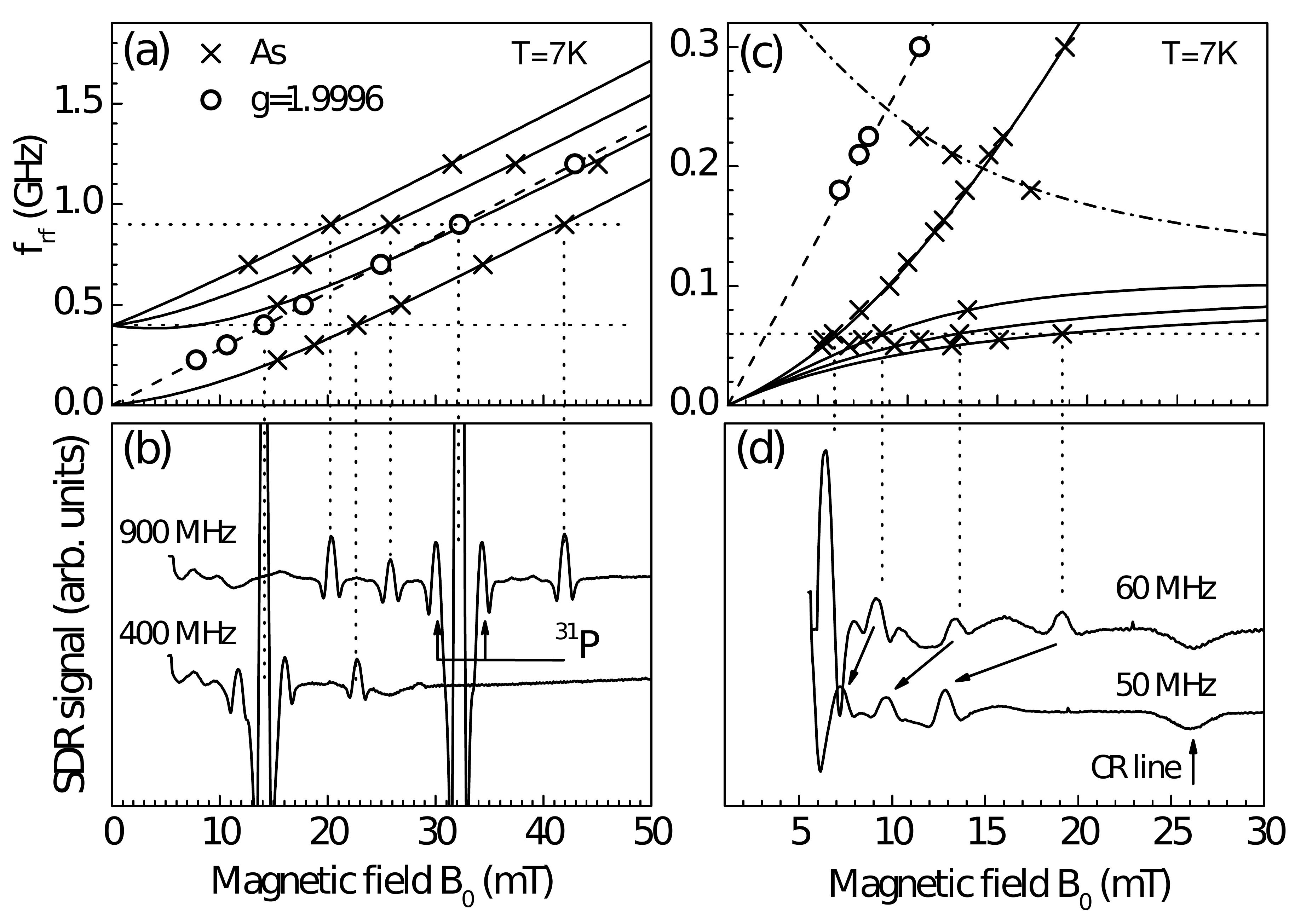}%
\caption{(a) Calculated resonance frequencies of the high-field-allowed $\Delta m_S=\pm 1$ As electron spin transitions (solid lines) and paramagnetic centers with $g=1.9996$ (dashed line).  Crosses and circles are the experimentally observed line positions for different resonance frequencies.  (b) SDR spectra of As-implanted FZ Si at $f_\mathrm{rf}=900$ and 400 MHz. (c) Calculated (lines) and experimental (crosses and circles) line positions for low-field transitions considering the mixing of electron and nuclear spin states ($\Delta m_F = \pm 1$ transitions). Full lines correspond to $\Delta F=0$, the dash-dotted line to $\Delta F =\pm 1$. (d) SDR spectra of As-implanted FZ Si detected at $f_\mathrm{rf}=60$ and $50$ MHz.}%
\label{figLowfield}%
\end{figure}

The spin Hamiltonian for a neutral As donor in a magnetic field $B_0$ (electron spin $S=1/2$ and nuclear spin $I=3/2$) is given by
\begin{align}
	\mathcal{H}_S=g\mu_e B_0 S_z-g_n \mu_n B_0  I_z+A \vec{S}\cdot\vec{I}\text{ ,}\label{EqH}
\end{align}
where $\mu_e$ and $\mu_n$ are the electron and nuclear Bohr magneton, respectively, $g = 1.99837$ and $g_n= 0.9596$  are the electron and nuclear $g$-factors, respectively, and $A/h= 198.35$ MHz is the hyperfine coupling \cite{feher_electron_1959}. For high magnetic fields ($g\mu_e B_z S_z \gg A \vec{S}\cdot\vec{I}$), four resonance lines corresponding to the allowed transitions with selection rules $\Delta m_S=\pm 1$, $\Delta m_I=0$ are observed in EPR, where $m_S$ and $m_I$ are the electron and nuclear spin projections, respectively. Calculated resonance frequencies $f_\mathrm{rf}$ for these transitions as a function of $B_0$ are shown by the solid lines in Fig.~\ref{figLowfield} (a). The observed line positions (crosses) for different radiofrequencies  $f_\mathrm{rf}$ are in very good agreement with the expected line positions. Figure \ref{figLowfield} (b) exemplary shows the SDR spectra observed at $f_\mathrm{rf}=900$ MHz and $400$ MHz. Together with the As resonance lines the additional signal corresponding to recombination centers with an isotropic $g$-factor $g=1.9996$ is observed, indicated in Fig.~\ref{figLowfield} (a) by the dashed line and the open circles. Two weaker lines originate from phosphorus donors present in the $n$-type silicon wafers.

The eigenstates of As donors at weak magnetic fields can be described by the total angular momentum $F=S+I$ ($F=1,2$) [cf.~Ref.~\onlinecite{wolfowicz_atomic_2013}] with the corresponding projections $m_F=-1,0,1$ and $m_F = -2,-1,\dots,2$, resulting in 6 allowed magnetic resonance transitions with  $\Delta F=0$, $\Delta m_F  = \pm 1$, two of which are degenerate [cf.~solid lines in Fig.~\ref{figLowfield} (c)]. Due to the very different coupling strengths of the driving magnetic field to the electron and nuclear spins, transitions corresponding to a change in $F$ can also be observed when $\Delta m_F=\pm 1$ [dash-dotted line in Fig.~\ref{figLowfield} (c)]. Indeed, the expected transitions are observed experimentally as a change of microwave photoconductivity in the SDR spectrum in Fig.~\ref{figLowfield} (d) and indicated by the crosses in Fig.~\ref{figLowfield} (c).

An additional line at $B_0 = 26$ mT is seen independently of the resonance frequency and is caused by the cross-relaxation (CR) taking place when the resonance frequencies of the paramagnetic centers with $g=1.9996$ and As coincide \cite{akhtar_electrical_2011, cochrane_zero-field_2012}.
Since the phase of the lock-in detection was chosen to maximize the As signal in the in-phase signal, the different dynamics of the CR process lead to a negative sign for this line. 


The observation of As and SL1 lines in the SDR spectra of the Cz-grown samples suggests the presence of an As--SL1 pair recombination process, which could offer a particularly sensitive spin readout as observed for $^{31}$P--SL1 pairs \cite{franke_spin-dependent_2014}. Therefore, pulsed high-field EDMR experiments were performed to further investigate the observed signal. 
To this end, a Cz-grown sample was prepared with an interdigit contact structure (Cr/Au) after the implantation (90 keV, $5\cdot 10^{14}$ cm$^{-2}$). The sample was neither annealed nor etched. It was placed in an X-band microwave resonator for pulsed EPR and a bias voltage of $U=0.5$ V was applied, resulting in a current of $I=10~ \mu$A under illumination with red light. 
These measurements were performed at a temperature of $T=8$ K. For pulsed EDMR experiments, short high-power microwave pulses are applied and the transient change in photocurrent is recorded directly after the last pulse. Integration of this transient gives a charge signal $\Delta Q$ proportional to the number of spin pairs in antiparallel configuration at the end of the pulse sequence \cite{boehme_theory_2003}.
 
\begin{figure}
\centering
\includegraphics[width=\linewidth]{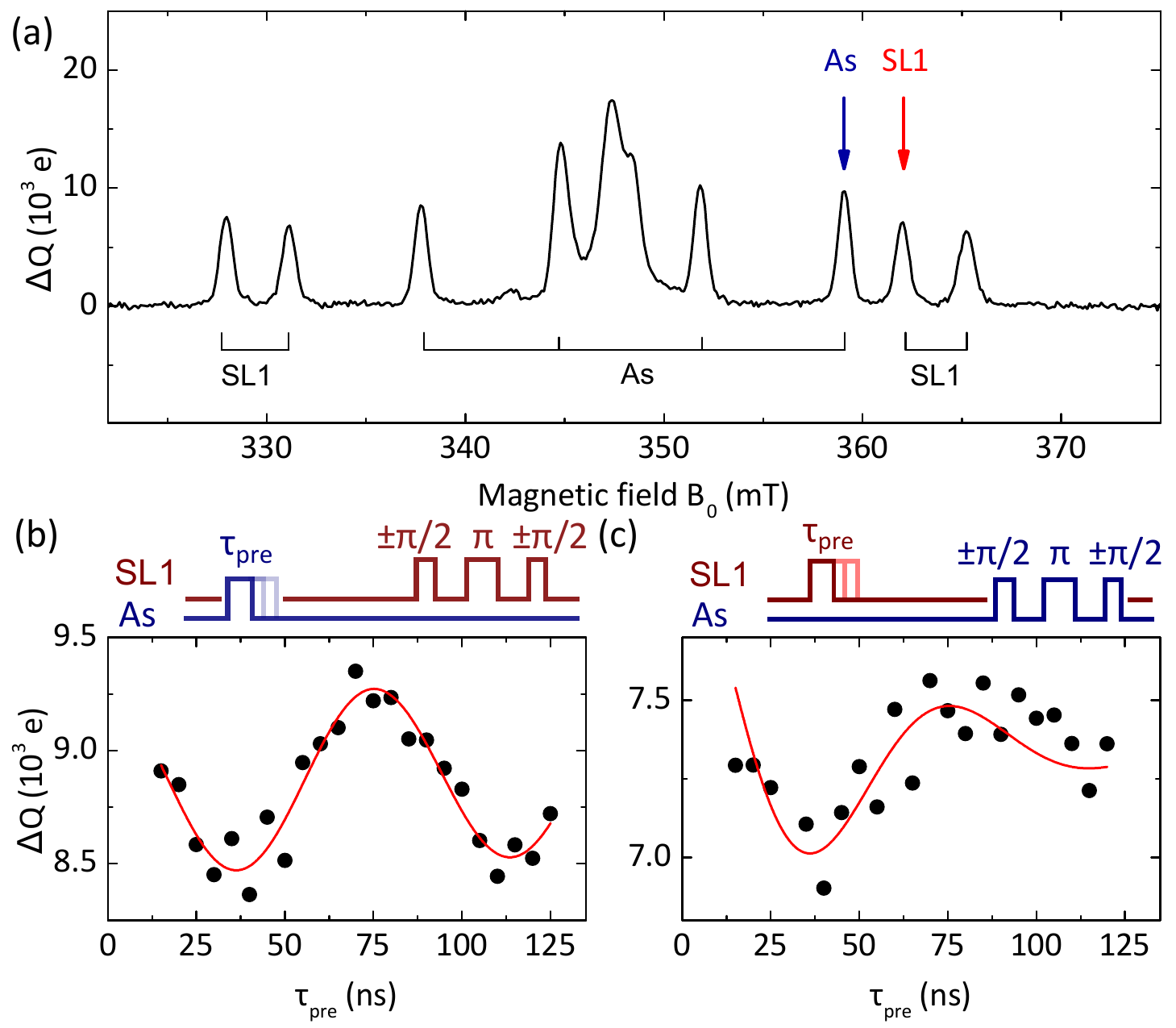}
\caption{(a) Echo-detected EDMR spectrum of As-implanted Cz-grown silicon for an angle of $\sim 10^{\circ}$ of the $\langle 111 \rangle$ crystal axis with the magnetic field $B_0$. (b) EDELDOR oscillations of the SL1 echo amplitude in dependence of the length $\tau_{\mathrm{pre}}$ of a preparation pulse on the As resonance. (c) EDELDOR oscillations of the As echo amplitude in dependence of the length $\tau_\mathrm{pre}$ of a preparation pulse on one of the SL1 resonances.}
\label{figPEDMR}
\end{figure}

Figure \ref{figPEDMR} shows $\Delta Q$ recorded after a spin-echo sequence with an extra $\pi/2$ pulse for the projection of the magnetization in the $x$-$y$-plane on the $z$ axis \cite{huebl_spin_2008} as a function of the magnetic field $B_0$. A four-step phase cycle was applied to implement a lock-in detection for an improved signal-to-noise ratio \cite{hoehne_lock-detection_2012}. As indicated in Fig.~\ref{figPEDMR}, the four allowed As transitions are observed as well as four resonances corresponding to different SL1 transitions \cite{brower_electron_1971}. Similar to experiments on $^{31}$P--SL1 spin pairs, the EDMR signal is observed as an increase in conductivity and can also be recorded after switching off the illumination (cf.~Ref.~\onlinecite{franke_spin-dependent_2014}).

To confirm the formation of As--SL1 pairs as the mechanism responsible for the observed EDMR signal, we use electrically detected electron double resonance (EDELDOR) \cite{hoehne_spin-dependent_2010}. In this pulsed EDMR method, pulses of two different frequencies are applied to probe the two partners of a pair process. 
Two frequencies $f_\mathrm{As}$ and $f_\mathrm{SL1}$ are chosen to match As and SL1 resonance lines [as schematically shown by the arrows in Fig.~\ref{figPEDMR} (a)], respectively, so that transitions for both spins can be excited at the same magnetic field $B_0$. A preparation pulse is applied on one of the resonances, followed by a detection echo sequence on the other resonance, as indicated by the pulse sequences in Fig.~\ref{figPEDMR} (b) and (c). Since for a pair recombination process $\Delta Q$ depends on the spin symmetry of the two partners, the detected signal will change with the length $\tau_\mathrm{pre}$ of the preparation pulse in this case. If, in contrast, the signals for SL1 and As result from independent mechanisms, a preparation pulse with frequency, e.g., $f_\mathrm{As}$ will have no influence on the detection echo with $f_\mathrm{SL1}$.
However, as shown in Fig.~\ref{figPEDMR} (b), an oscillation of the SL1 signal amplitude is observed as a function of $\tau_\mathrm{pre}$, reflecting the Rabi oscillations induced by the preparation pulse on the As resonance. Since only one of the four As lines is addressed and the line width [$\sim 25$ MHz, cf.~Fig.~\ref{figPEDMR} (a)] is larger then the excitation bandwidth of the microwave pulses [$2\pi$ pulse length $\tau_{2\pi}\approx 75~\mathrm{ns}=1/(13.3$ MHz)], an oscillation depth of about 15 \% is expected, which is in good agreement with experiment ($\sim 12$\%). The EDELDOR experiment was repeated with the roles of the As and SL1 resonances interchanged. Here, an oscillation of the As signal amplitude could be observed corresponding to the Rabi oscillations driven by the preparation pulse at frequency $f_\mathrm{SL1}$ [Fig.~\ref{figPEDMR} (c)]. In this case, the relative oscillation depth is slightly smaller, which is consistent with the broader linewidth of the SL1 peak in the spectrum [Fig.~\ref{figPEDMR} (a)]. The time constants for the recombination process were determined to be $\tau_\mathrm{p}=610(10)~\mu$s  and $\tau_\mathrm{ap}=4.6(5)~\mu$s for the parallel and antiparallel recombination time, respectively, following the approach summarized in Ref.~\onlinecite{franke_spin-dependent_2014}. These values are of the same order of magnitude as the time constants observed for corresponding $^{31}$P--SL1 pairs.

In conclusion, we have demonstrated that SDC can be used for the investigation of paramagnetic centers produced in low dose low energy implantation in Si and experimental conditions for the observation of isolated As donors and radiation defects were found. Using SDR at low magnetic fields, resonance transitions between As spin states corresponding to a change $\Delta m_F=\pm 1$ could be detected, some of which are forbidden in the high-field limit. Pulsed EDMR experiments on As-implanted Cz Si show that spin pairs of As donors and SL1 centers are formed and can provide a sensitive readout of the donor spin state. Furthermore, the time constants of the involved spin-dependent process were determined allowing to realize advanced EDMR measurements like pulsed electron nuclear double resonance (ENDOR) of the $^{75}$As nuclear spin on these types of samples.

This work was supported in part by a Project for Developing Innovation Systems by MEXT, JSPS Core-to-Core, and DFG (Grant No.~SFB 631, C3 and Grant No.~Br 1585/5).


\bibliography{manabu}

\end{document}